\documentclass[5p,twocolumn,times,nonatbib]{elsarticle}
\usepackage{graphicx}
\usepackage{amsmath}   
\usepackage{etoolbox}
\usepackage{amssymb}
\usepackage{epsfig}
\usepackage{lineno, hyperref}
\usepackage{url}
\usepackage[export]{adjustbox}
\usepackage[font=footnotesize,labelfont=bf,skip=5pt]{caption}
\usepackage{subcaption}
\usepackage{xpatch}
\usepackage{xspace}
\usepackage{float}
\makeatletter
\let\c@author\relax
\makeatother
\usepackage[backend=biber,style=phys,eprint=true,hyperref=true,biblabel=brackets,block=ragged]
{biblatex}

\usepackage{hyperref}
\hypersetup{
    colorlinks=true,
    linkcolor=blue,
    filecolor=magenta,      
    urlcolor=cyan,
    pdftitle={proANUBIS Construction},
    pdfpagemode=FullScreen,
    }
\def\pmbanner{}

\newcommand{\eg}{\mbox{\itshape e.g.}\xspace}
\newcommand{\ie}{\mbox{\itshape i.e.}\xspace}

\newcommand{\metre}{\ensuremath{\textnormal{m}}\xspace}

\def\ns{\text{ns}\xspace}

\def\metre{\text{m}\xspace}

\newcommand{\proanubis}{\mbox{\textsl{pro}ANUBIS}\xspace}
\newcommand{\atlas}{\mbox{ATLAS}\xspace}

\begin{document}
\begin{frontmatter}
\title{\pmbanner \LARGE Commissioning of \proanubis:\\ A proof-of-concept detector for the ANUBIS experiment\\
\includegraphics[width=0.1\linewidth]{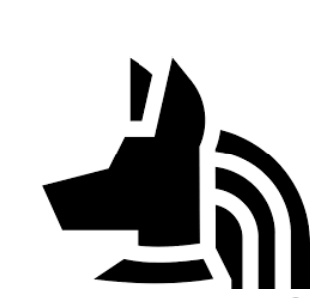}
\\
{\large\textbf{ANUBIS Collaboration}}
\\[-0.2em]
{\normalsize\textit{E-mail: }\href{mailto:anubis-publications@cern.ch}{anubis-publications@cern.ch}}
\\[-0.2em]}
\author[18]{Giulio Aielli}
\author[1]{Oleg Brandt}
\author[20]{Patrick Collins}
\author[3]{Louie Dartmoor Corpe}
\author[1]{Jonas Dej}
\author[17]{Oliver Kortner}
\author[17]{Hubert Kroha}
\author[1]{Christopher Lester}
\author[9]{Luca Pizzimento}
\author[2]{Ludovico Pontecorvo}
\author[1]{Michael Revering}
\author[1]{Aashaq Shah}
\author[17]{Daniel Soyk}
\author[1]{Paul Swallow}
\author[6,20]{Yanglin Wan}
\author[]{for the ANUBIS Collaboration}

\address[1]{Cavendish Laboratory, University of Cambridge, Cambridge, United Kingdom}
\address[2]{CERN, Geneva, Switzerland}
\address[3]{Universit\'e Clermont Auvergne / Laboratoire de Physique de Clermont Auvergne, CNRS/IN2P3, France}
\address[9]{Department of Physics, University of Hong Kong, Hong Kong SAR, China}
\address[17]{Max Planck Institute for Physics, Munich, Germany}
\address[18]{University and INFN Roma Tor Vergata, Rome, Italy}
\address[20]{Formerly at: Cavendish Laboratory, University of Cambridge, Cambridge, United Kingdom}
\vspace{-0.5cm}
\begin{abstract}
Long-lived particles (LLPs), predicted by various extensions of the Standard Model (SM), have become a key focus of the contemporary search programme for physics beyond the SM. 
To enhance LLP discovery potential at the LHC, the ANUBIS experiment has been proposed to instrument the ceiling of the \atlas experiment's underground cavern with dedicated tracking detectors. This report summarises recent progress towards realising ANUBIS.

Specifically, a key milestone has been achieved with the installation and commissioning of \proanubis, a prototype that serves as a proof-of-concept for ANUBIS. We describe the \proanubis setup, including its remotely-operated data acquisition system and automatic signal processing chain. 
The \proanubis demonstrator 
is used to evaluate the detector performance under realistic conditions in the UX1 \atlas experimental cavern, including readout synchronisation with the \atlas experiment. 
Furthermore, \proanubis allows for the direct measurement of relevant background processes in a representative location within the \atlas cavern, providing input for the simulation of such processes for the future ANUBIS detector. 
The paper concludes with an update on the current status of the ANUBIS project and its roadmap toward a full-scale implementation in the \atlas cavern.
\begin{center}
    \today 
\end{center}
\end{abstract}

\end{frontmatter}
 
\section{Introduction} 
\label{Sec:Intro}
Long-lived particles (LLPs), \ie particles with macroscopic lifetimes $\tau>10$~ps, emerge in a multitude of Beyond the Standard Model~(BSM) theories~\cite{Alimena:2019zri}, offering potential explanations for fundamental mysteries such as 
dark matter~\cite{Hall:2010jx,Cheung:2010gk,Zurek:2013wia}, 
neutrino masses~\cite{Bondarenko:2018ptm}, 
neutral naturalness~\cite{Giudice:1998bp,Burdman:2006tz}
and the baryon asymmetry of the Universe~\cite{Canetti:2012kh,Asaka:2005pn}. 
Their lifetimes can range from 10~ps up to the Big Bang Nucleosynthesis limit~\cite{Fradette:2017sdd, Kahniashvili_2022}.
However, neutral LLPs with longer lifetimes present a unique challenge for detection, as the sensitivity of the main detectors of the Large Hadron Collider (LHC) at CERN is hampered by intricate backgrounds, constraints from trigger hardware and bandwidth, and limited acceptance due to their finite volume close to the beamline~\cite{CMS:2021sch, CMS:2021juv}. 
To explore the domain of long lifetimes resulting in decay lengths of $\mathcal{O}(10~\metre)$ and above, several proposals with dedicated detectors have recently emerged, intending to search for LLPs with moderate to relatively long lifetimes. Notable among these initiatives are AL3X~\cite{Gligorov:2018vkc}, ANUBIS~\cite{Bauer:2019vqk}, \mbox{CODEX-b}~\cite{Gligorov:2017nwh}, FASER and FASER2~\cite{Feng:2017uoz, FASER:2018eoc},  MATHUSLA~\cite{aitken2025conceptualmathusla, Chou:2016lxi, Curtin:2018mvb, MATHUSLA:2020uve}, MoEDAL-MAPP1 and MoEDAL-MAPP2 of the MoEDAL collaboration~\cite{Pinfold:1181486, Pinfold:2019zwp}, and SHiP~\cite{Bonivento:2013jag}. 
This paper focuses on the ANUBIS project, particularly on its proof-of-concept prototype detector, \proanubis, that has been installed in the UX1 underground experimental cavern of the \atlas experiment at Interaction Point 1 (IP1) of the LHC~\cite{Aad:1129811}.

The report is organised as follows:
Section~\ref{Sec:ANUBIS_overview} briefly introduces the ANUBIS project and highlights its potential to search for LLPs. 
Section~\ref{Sec:proANUBIS_demon} presents the motivations of the \proanubis detector, Section~\ref{Sec:Commision_Install} details its installation at IP1 of the LHC and commissioning,  Section~\ref{Sec:Signal_processing} covers its signal processing and data acquisition system, and
Section~\ref{Sec:Remote_Control} discusses its control system. 
Section~\ref{Sec:Ambient_monitoring} outlines the automated monitoring system of the ambient conditions relevant for the
\proanubis detector. 
Section~\ref{sec:CommissioningData} highlights the initial results obtained from operating \proanubis during commissioning. 
Finally, Section~\ref{Sec:Summary} provides an overview of more recent efforts and outlines future plans.

\section{Overview of the ANUBIS Experiment} 
\label{Sec:ANUBIS_overview}

The ANUBIS experiment~\cite{Bauer:2019vqk} is an LLP detector proposed for installation in the UX1 experimental cavern of the \atlas detector. 
The primary physics target of the ANUBIS experiment is to explore the landscape of LLPs produced at partonic centre-of-mass energies corresponding to the electroweak scale or above, thereby complementing the sensitivity of FASER~\cite{Feng:2017uoz}, SHiP~\cite{Bonivento:2013jag,SHiP:2021nfo}, and main LHC detectors like ATLAS, CMS, and LHCb.
The key idea behind the ANUBIS detector concept is utilising the air-filled volume extending to the \atlas detector while minimising civil engineering requirements by repurposing the existing infrastructure of the \atlas experiment~\cite{Aad:1129811} at the LHC.

Simulation studies have been used to evaluate various geometrical layouts for ANUBIS, including the deployment of tracking layers within the main \atlas service shaft PX14, instrumenting the UX1 cavern ceiling, and hybrid configurations combining both options. Among these, the instrumentation of the cavern ceiling and the bottom of the PX14 and PX16 service shafts has emerged as the most promising design~\cite{ANUBIS:2025sgg}, offering significantly enhanced sensitivity to neutral LLP decays compared to existing or approved LHC detector systems, while providing similar or superior performance to other LLP proposals~\cite{PBC:2025sny}.  

\begin{figure}[ht]
\centering
\includegraphics[height=7.6cm]{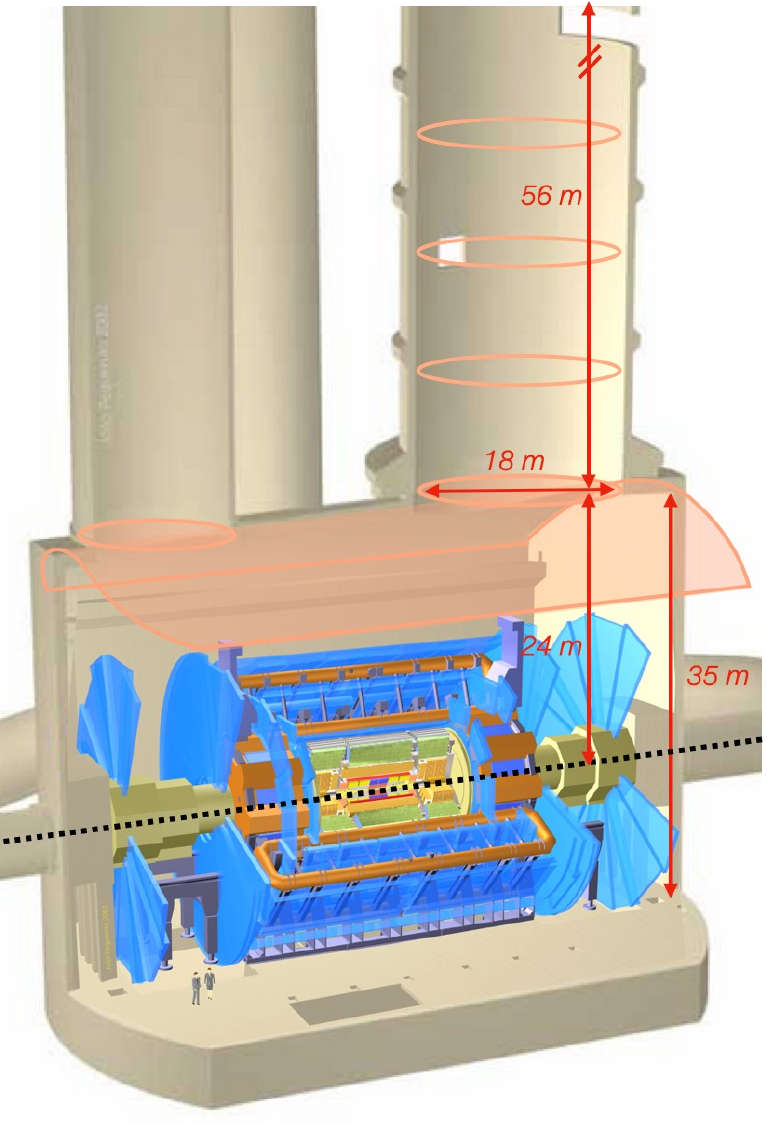} 
\caption{
Sketch of the layout of the underground cavern at IP1 of the LHC, featuring the \atlas experiment, and the PX14 and PX16 access shafts. 
The area highlighted in orange illustrates the potential ceiling configuration of the ANUBIS detector. 
In addition to the instrumented cavern ceiling, this configuration includes two discs covering the access shafts.
} 
\label{fig:SketchATLAS_undergroundCavern}
\end{figure}

The baseline ANUBIS configuration, referred to as the {`ceiling configuration'} in the following, envisions instrumenting tracking detectors on the ceiling of the \atlas underground cavern in two detector layers separated by $\sim1~\metre$, together with an additional disc with the same two-layer layout of tracking detectors installed in each of the PX14 and PX16 service shafts, as illustrated in Figure~\ref{fig:SketchATLAS_undergroundCavern}~\cite{ANUBIS:2025sgg}. 
This configuration extends the discovery reach of \atlas to LLP decays with vertices occurring between the primary detector volume and the cavern ceiling, thereby targeting a region of LLP parameter space that is not covered by the existing or approved detectors at the LHC.

\section{The \proanubis Detector} 
\label{Sec:proANUBIS_demon}
The ANUBIS concept envisaged the installation of a prototype detector, \proanubis, to provide a proof-of-concept demonstrating the feasibility of the experiment, and to characterise the background environment relevant for the LLP search program with the ANUBIS detector. 
In addition, the \proanubis detector was expected to provide first-hand operational experience under the radiation and environmental conditions of the \atlas underground cavern, as well as to support the development of institutional expertise and collaboration across the project.

\subsection{Detector Design and Geometry}
\label{sec:proANUBISDetector}

The \proanubis detector consists of three tracking layers of BIS7-type Resistive Plate Chambers (RPCs)~\cite{Massa:2020hjw}, as detailed in Ref.~\cite{ANUBIS:2025mme}. 
BIS7 is a technology that is similar to what would be used for ANUBIS.
It is very similar to the \atlas Phase-II RPCs~\cite{CERN-LHCC-2017-020}, with the main notable difference in the readout geometry. 
The three layers of \proanubis are arranged vertically with an inter-layer spacing of $0.6~\metre$ ($0.5~\metre$) between the bottom (top) two tracking layers, providing a total lever arm of $1.1~\metre$.

\begin{figure}[ht]
\centering
\begin{subfigure}{\linewidth}
  \centering
  \includegraphics[width=6.5cm,height=5cm]{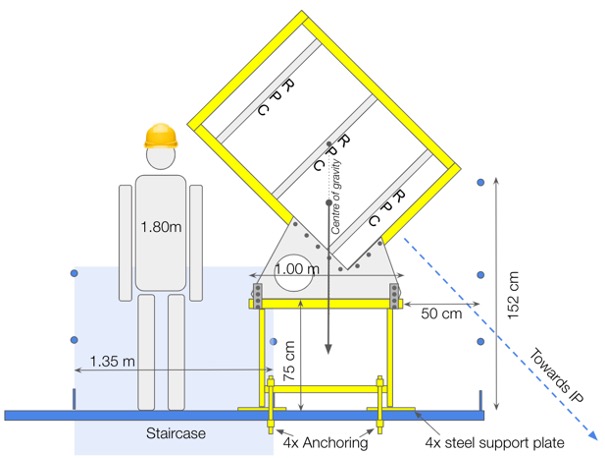}
  \caption{}
  \label{fig:proANUBIS_installation_position_Top}
\end{subfigure}
\begin{subfigure}{\linewidth}
  \centering
  \includegraphics[width=6.5cm,height=4.5cm]{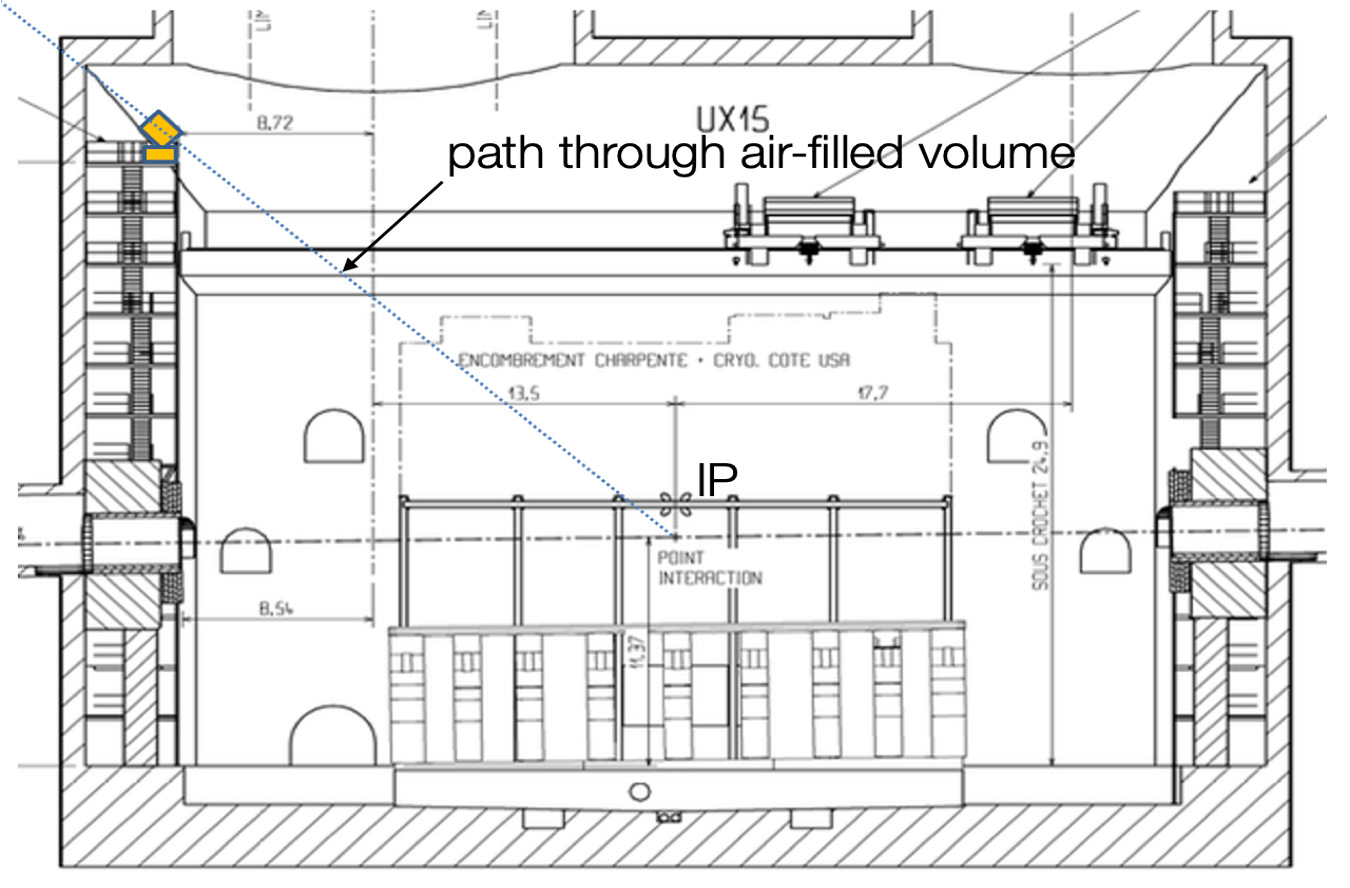}
  \caption{}
  \label{fig:proANUBIS_installation_position_Low}
\end{subfigure}
\caption{(a) Sketch showing the design of \proanubis metallic frame (in yellow) and its support structure. 
(b) The location and orientation of the \proanubis setup within the UX1 \atlas cavern.}
\label{fig:proANUBIS_installation_position}
\end{figure}

The bottom layer of \proanubis corresponds to one RPC module composed of three RPC detectors stacked to form a `triplet' within a mechanical frame, while the top layer module contains two RPCs forming a `doublet', and the middle layer module features a single RPC detector referred to as a `singlet'. 
Each RPC detector has an active area of approximately \(1~\metre \times 1.8~\metre\), where the short side corresponds to the \(\eta\)-readout plane with readout strips parallel to the long side, and vice versa for the orthogonal $\phi\text{-readout}$ plane\footnote{This naming convention is adopted from the \atlas BIS7 project where the RPCs measure the $\eta$ and $\phi$ coordinates of a hit.}. 
Hence, the two planes provide a two-dimensional spatial measurement. 
The design of the assembled \proanubis detector is shown in Figure~\ref{fig:proANUBIS_installation_position}, where it is angled to maximise the geometrical acceptance to particles originating from the interaction point (IP). 
The installation site is located on Level~12 of Side~A inside the \atlas cavern, approximately 80~m below the Earth's surface and close to the ceiling of the UX1 \atlas cavern was chosen to replicate the expected environment of ANUBIS, as shown in the same Figure~\ref{fig:proANUBIS_installation_position}. 

\section{Installation and Commissioning of \proanubis} 
\label{Sec:Commision_Install}
\begin{figure}[ht]
\centering
\begin{subfigure}{\linewidth}
\centering
\includegraphics[width=5cm, height=5.7cm]{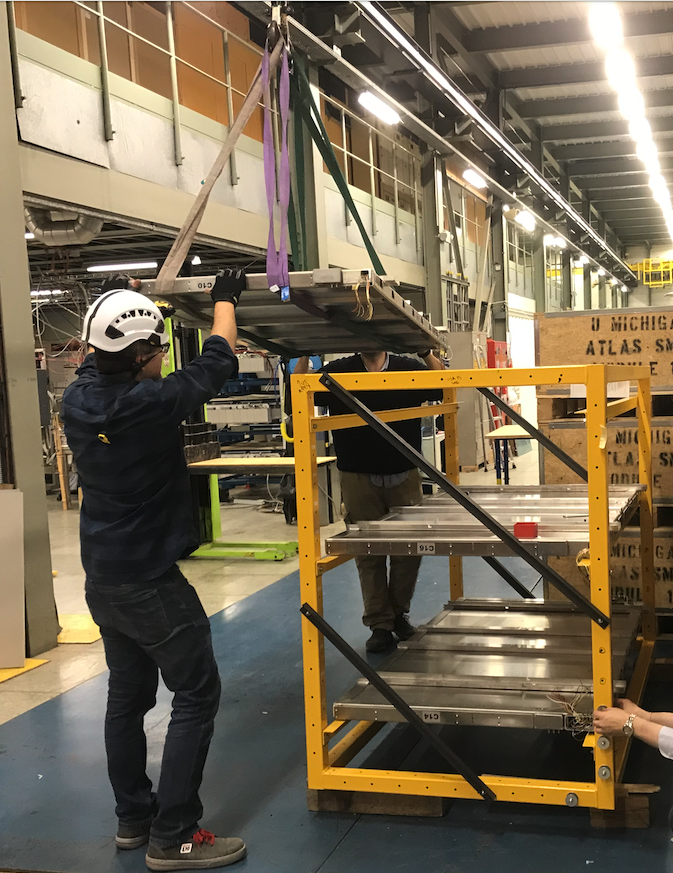} 
  \caption{}
  \label{fig:proANUBIS_loading_transportation_Top}
\end{subfigure}
\begin{subfigure}{\linewidth}
\centering
\includegraphics[width=6.4cm, height=4.4cm]{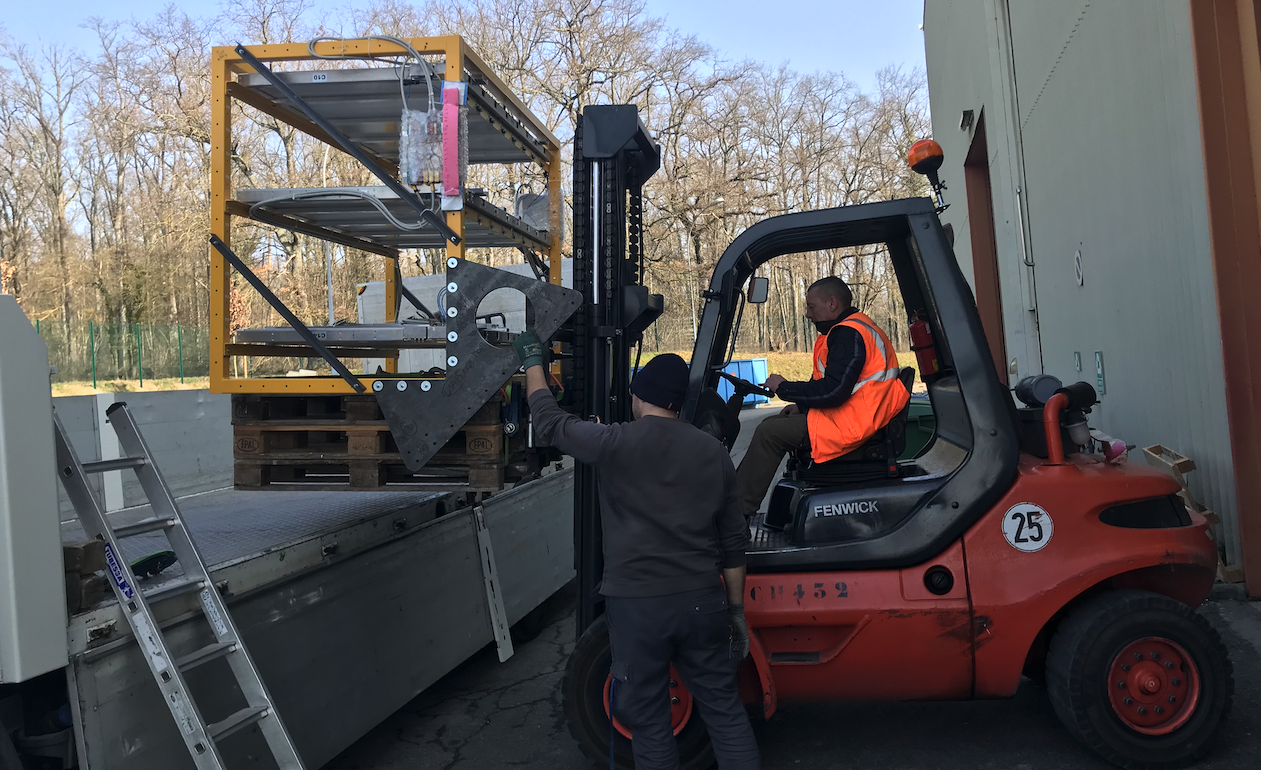} 
  \caption{}
  \label{fig:proANUBIS_loading_transportation_Low}
\end{subfigure}
\caption{
(a) The yellow \proanubis steel frame being populated with RPC integrated chambers at CERN's BB5 laboratory.
(b) \proanubis after integration of on-detector services being lifted at BB5 for transport to IP1.} 
\label{fig:proANUBIS_loading_transportation}
\end{figure}

The commissioning of \proanubis commenced at CERN’s BB5 laboratory, where the integration and initial performance tests were carried out~\cite{ANUBIS:2025mme}.
The triplet, singlet, and doublet RPC modules were installed into the steel support frame, 
as illustrated in Figure~\ref{fig:proANUBIS_loading_transportation_Top}, in accordance with the mechanical design specifications.
In the next step, all on-detector services were integrated, and the high- and low-voltage~(HV and LV) systems were tested at CERN's BB5 laboratory to verify operational functionality. 
In addition, preliminary tests of the data acquisition system described in Section~\ref{sec:proANUBIS-DAQ0} were carried out. 
Once all components had been validated and qualified for deployment, the complete \proanubis assembly was transported (Figure~\ref{fig:proANUBIS_loading_transportation_Low}) to the SX1 hall above IP1 in preparation for deployment in the UX1 experimental cavern of the ATLAS detector.

\begin{figure}[ht]
\centering
\begin{subfigure}{\linewidth}
\centering
\includegraphics[width=7cm, height=5cm]{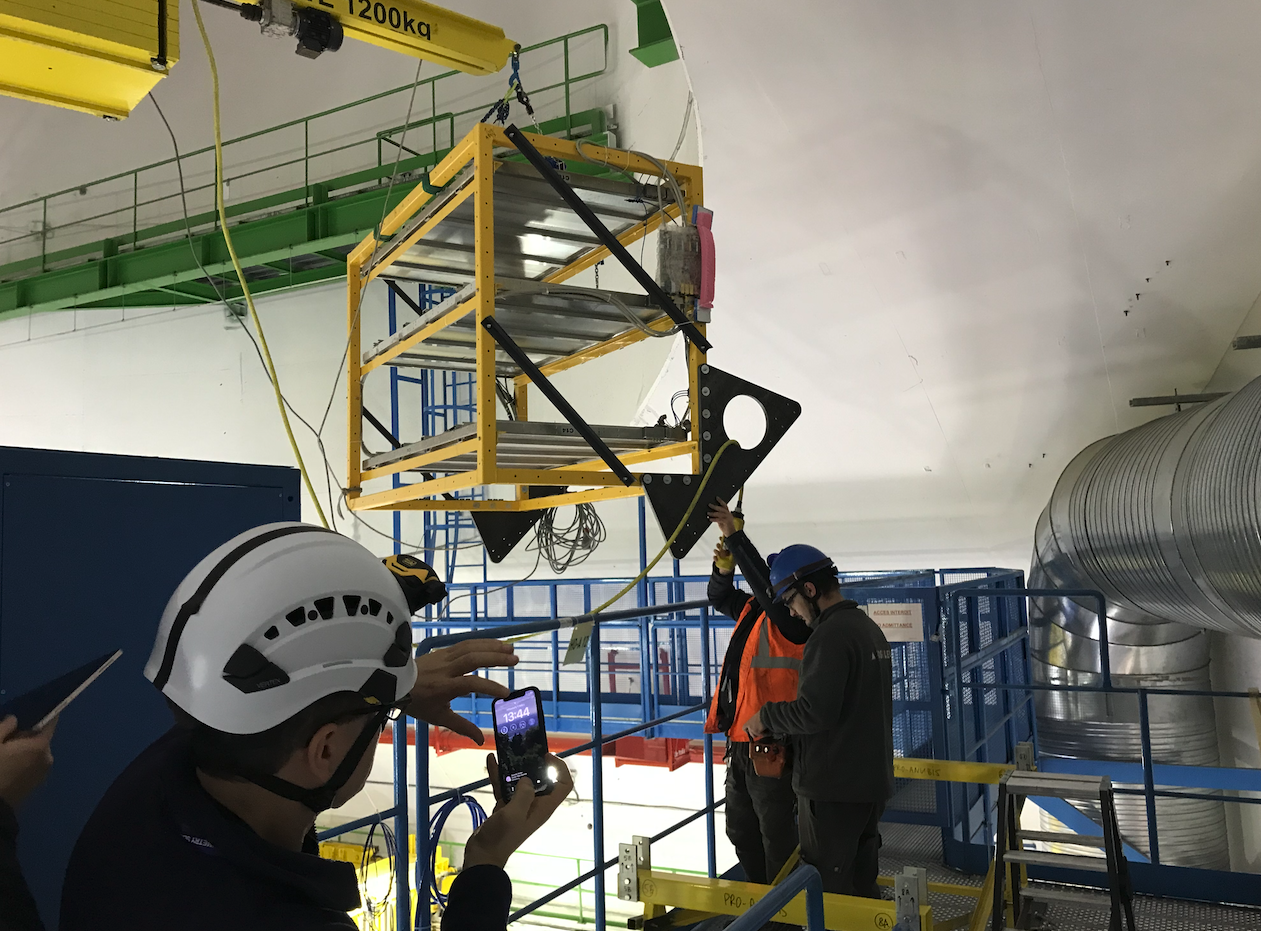} 
\caption{}
  \label{fig:proANUBIS_loading_transportation_instalation_Top}
\end{subfigure}
\begin{subfigure}{\linewidth}
\centering
\includegraphics[width=7cm, height=5cm]{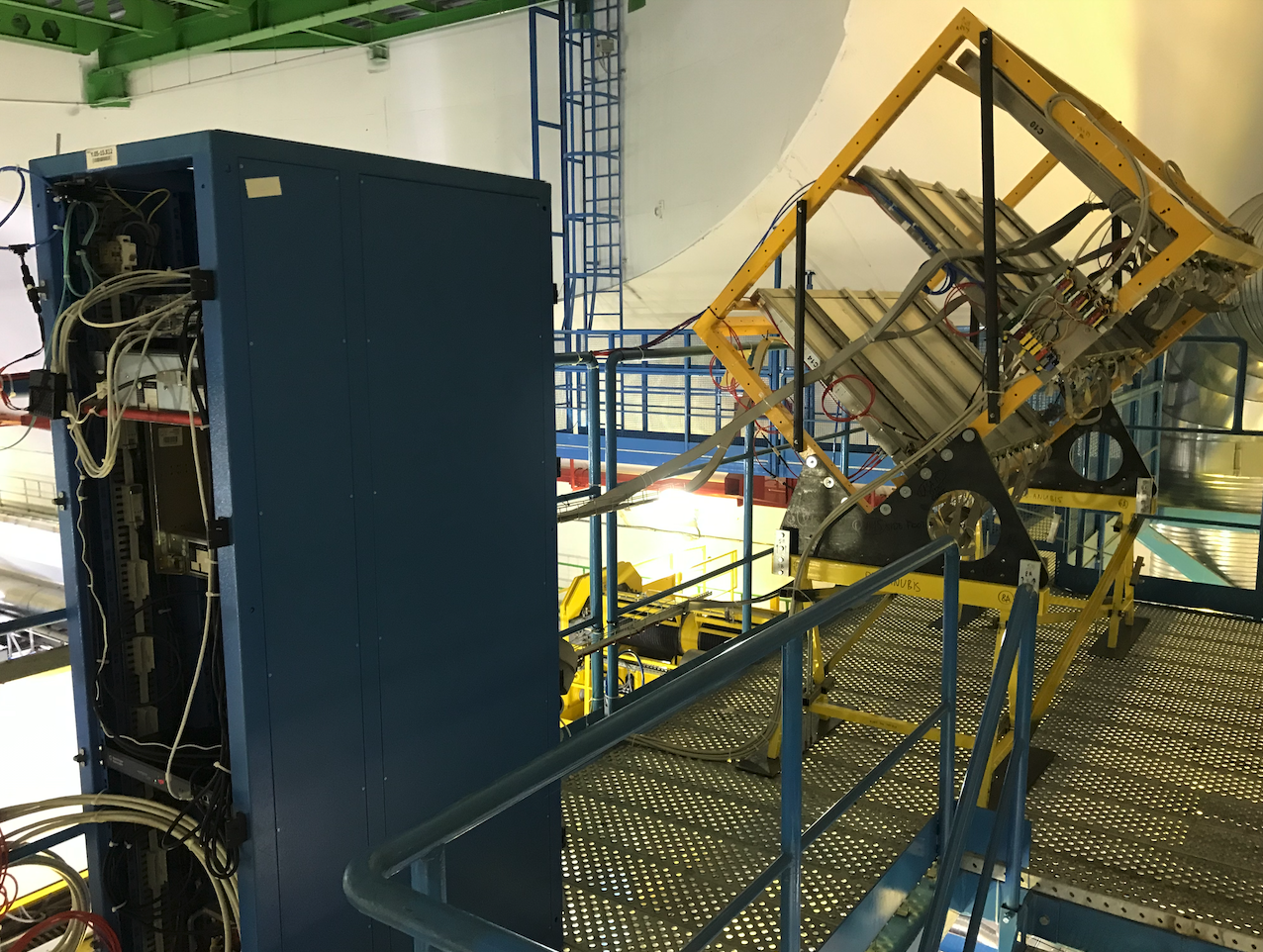} 
\caption{}
  \label{fig:proANUBIS_loading_transportation_instalation_Low}
\end{subfigure}
\caption{
(a) \proanubis being lowered with the aid of a crane through the \atlas PX14 access shaft.  
(b) The final installation of \proanubis, along with the primary DCS and DAQ rack, on Level 12 Side A in the UX1 ATLAS cavern.
The detector panels of \proanubis are approximately perpendicular to the line-of-sight towards the IP, which is roughly towards the bottom left corner of the picture.
} 
\label{fig:proANUBIS_loading_transportation_instalation}
\end{figure}

With the support of the \atlas Technical Coordination team, the \proanubis detector, together with the primary Detector Control System (DCS) and Data AQuisition (DAQ) rack~(Section~\ref{sec:proANUBIS-DAQ0}), was lowered into its final position through the PX14 access shaft using a crane system for precise manoeuvring, as shown in Figure~\ref{fig:proANUBIS_loading_transportation_instalation_Top}.
The coordinated efforts helped the successful installation of the \proanubis detector and its associated DCS and DAQ infrastructure within the \atlas experimental cavern, as illustrated in Figure~\ref{fig:proANUBIS_loading_transportation_instalation_Low}~\cite{Shah:2024fpl}. 
All installation activities were completed during the Year-End Technical Stop (YETS) of 2022, ensuring that the process caused no interference with ongoing LHC operations.

\begin{figure*}[t] 
\centering
\includegraphics[width=15cm]{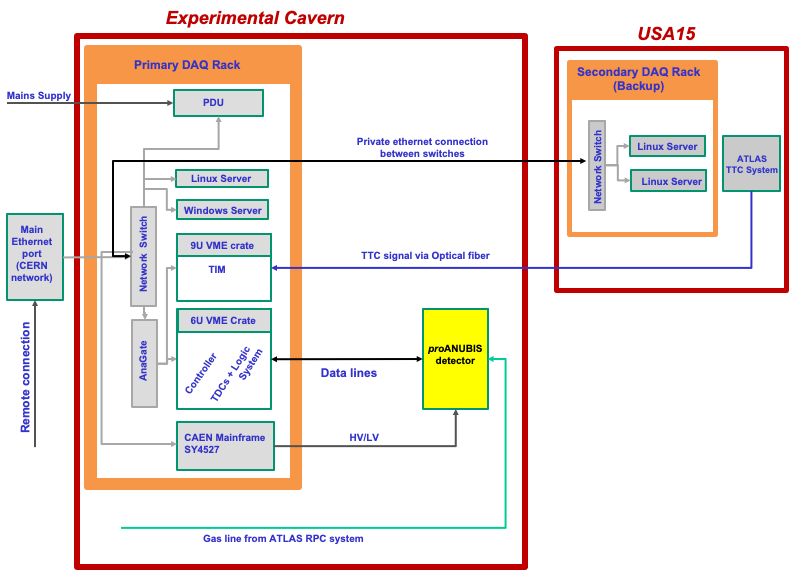}
\caption{
Schematic layout of the \proanubis experimental setup. 
The system comprises two interconnected parts: the underground UX1 ATLAS experimental cavern hosting the \proanubis detector and the primary DCS and DAQ rack, and the USA15 area hosting the secondary DAQ rack.
The primary DCS and DAQ rack houses the CAEN SY4527 mainframe hosting LV and HV power supplies, the VME crates with a VME controller, TDCs, hardware trigger logic boards, and the timing, trigger, and control interface module for receiving the ATLAS central trigger processor clock. 
The secondary DAQ rack houses two servers providing backup and full operational redundancy in case of a primary server failure in the cavern.
The experimental cavern and USA15 are linked via network switches operating on a dedicated private network to ensure continuous communication and remote control capability.
}
\label{fig:proANUBIS_sketch_plus_DAQ}
\end{figure*}

A schematic overview of the fully installed detector and its associated DCS and DAQ systems is shown in Figure~\ref{fig:proANUBIS_sketch_plus_DAQ}. 
The experimental equipment is installed in two locations: Level~12 Side~A of the UX1 cavern labelled  `Experimental cavern' that hosts the \proanubis detector together with the DCS and DAQ rack, and the adjacent radiation-protected USA15 cavern that accommodates a secondary DAQ rack equipped with backup servers.
The USA15 section also hosts the interface to the \atlas Timing, Trigger, and Control (TTC) system, which distributes the clock signal from the Central Trigger Processor that is interfaced to the global LHC clock and ATLAS Level-1 trigger information to \proanubis. 
These signals, discussed in a later section, enable the synchronisation of the \proanubis and ATLAS DAQ systems, allowing the correlation of events recorded independently by the \proanubis and \atlas detectors.

\subsection{Gas System}
\label{sec:GasSystem}
The gas supply for \proanubis is provided through the \atlas Muon RPC gas system, operating using the same gas mixture as the \atlas RPC detectors, allowing \proanubis to automatically follow any changes or calibrations applied to the main \atlas gas infrastructure.
The system allows a continuous and stable supply of the standard ATLAS RPC gas mixture at a flow rate of approximately 5--6~l/h. 
The standard ATLAS RPC gas mixture comprises 94.7$\%$ C$_{2}$H$_{2}$F$_{4}$ (Freon, commercially known as R134a), 5$\%$ i-C$_{4}$H$_{10}$ (isobutane), and 0.3$\%$ SF$_{6}$. 
In 2024, the gas mixture was changed to include an additional CO${_2}$ component of 30\%, with the other component gases roughly at the same ratios of the standard gas mixture, except for SF$_{6}$, which was increased to 1\%.

\subsection{Primary DCS and DAQ rack} 
\label{sec:proANUBIS-DAQ0}
The primary DCS and DAQ rack is located at Level 12 Side A within the UX1 experimental cavern and hosts the key elements required for detector control and data acquisition. 
These include a Power Distribution Unit (PDU); 
two independent computer servers; 
a 9U VME crate housing the Timing, trigger, and control Interface Module (TIM)~\cite{Postranecky:529414}; 
and a 6U VME crate housing components such as Time-to-Digital Converters (TDCs), a VME controller, and trigger logic boards. 
In addition, the CAEN SY4527 mainframe is installed in the rack with HV and LV power supplies serving the detector and associated electronics.

\subsubsection{Computer Servers}
There are two primary computer servers, one with a Windows Server 2016 and one with an Alma Linux 9 Operating System. 
Each of them can control the data acquisition from the \proanubis detector as well as monitor and control its operating conditions. 
Depending on software compatibility and operational requirements, either server can be selected to control the \proanubis DCS and DAQ systems, ensuring continuous and reliable operation of all hardware components throughout data taking. 
In addition to the two primary servers, there are also two backup computer servers located in USA15 to provide operational redundancy in case of hardware failure in the cavern, see Section~\ref{proAnubis-DAQ1}.

\subsubsection{HV/LV Control System} 
The CAEN SY4527 mainframe provides power to the detector and associated electronics. 
It is equipped with a primary HV supply unit (CAEN model A4531), a 12-channel HV module (model A1832P), and three LV supply modules (models A1517B/A1518A), offering a total of 18 LV channels. 
The HV system delivers the required operating voltage to the RPC gas gaps, with six HV channels powering the six RPCs of \proanubis. 
The LV system supplies power to the Front-End (FE) electronics, with 12 FE boards per chamber, for a total of 72 FE boards across the detector.
The LV are provided independently for the $\eta$ and $\phi$ side for flexibility.
The discriminator threshold voltage reference is additionally divided up by RPC module (top/middle/bottom) to allow studies of its effect on the detector performance.

\subsubsection{CAEN VME-Ethernet controller bridge} 
\label{sec:CAENcontroller}
A CAEN VME-Ethernet controller bridge (model V4718), also referred to as the `controller' in this report, is used to interface with and control the TDCs housed in the VME crate. 
The controller is accessible from any of the four computer servers, either the main two in the experimental cavern or the backup servers. 

\subsubsection{Time-to-Digital Converters} 
\label{sec:CAENTDCs}
The CAEN V767 TDC modules form the core of the DAQ readout system. 
They digitise and record the arrival times of signals from the RPC detectors, providing the timing information of hits required for event reconstruction and offline analysis. 
The TDCs feature a time binning of 0.8~ns, which is adequate for the performance studies targeted by \proanubis and are broadly compatible with the sub-nanosecond intrinsic timing performance of the RPC detector technology. 

The installed configuration employs five TDC modules to read out a total of 576 detector channels (six identical RPC detectors with 96 channels each) when they receive an external trigger signal. 
An additional TDC is dedicated to recording trigger and LHC clock information distributed through the TIM~(Section~\ref{SubSec:TIM}), enabling synchronisation of the \proanubis and \atlas detectors and hence studies of events recorded by both detectors.

During YETS 2024--25 the DAQ system was upgraded to use CAEN V1190a TDCs that provide a time binning of 100~ps.

\subsubsection{Timing Trigger and Control Interface Module} 
\label{SubSec:TIM}
The TIM links the \proanubis DAQ system to the \atlas Level-1 Trigger infrastructure~\cite{Postranecky:529414}. 
It receives the global TTC stream from \atlas via an optical fibre to a TTC receiver chip (TTCrx), decodes the information, and distributes selected signals to the \proanubis DAQ system. 
The 40~MHz LHC clock is delivered from the TIM to the TDCs as a dedicated \emph{control} input in the Emitter-Coupled Logic (ECL) format, providing the timing reference for digitisation. 
The Level-1 Accept (L1A), Bunch Crossing Reset (BCR), and Event Counter Reset (ECR) signals are injected into the TDC as Low-Voltage Differential Signaling (LVDS) data lines and recorded using a dedicated TDC in the same manner as ordinary detector signals; their semantics are reconstructed in software during event building to ensure correct synchronisation and event identification within the \proanubis data stream.

A fixed TTC signal distribution latency of up to approximately \(3~\mu\text{s}\) arises from the cabling path between the \atlas experiment and the \proanubis detector location in the cavern. This latency is stable and deterministic, and it is accounted for in the readout configuration.

\subsubsection{Trigger OR Logic Board} 
\label{SubSec:TOR}
The Trigger OR (TOR) board is a custom-designed VME-style module used to derive a trigger signal from the RPC detector hit information while preserving the original data for further readout using the TDCs. 
It accepts up to 32 LVDS inputs through a single Robinson Nugent connector. Each input signal is first buffered and split into two identical paths. 
One path is routed to a 32-channel LVDS Robinson Nugent output connector that provides the readout signal to the TDCs, while the second path is combined through a logical OR network to generate a trigger signal whenever at least one input channel registers a hit.
The data flow path is schematically represented in Figure~\ref{fig:proANUBIS_signal_processing}.

The trigger output generated by the TOR board is formatted in Transistor-Transistor Logic (TTL) and features an adjustable gate width of 20--100~ns (operated at 60~ns in 2024 and 2025 data taking), with an internal processing latency of the board of $\sim10\ns$. 
The TTL trigger output signal is made available on a LEMO connector for routing to the remainder of the \proanubis DAQ chain. 

\subsubsection{Majority Logic OR Board} 
\label{SubSec:SOR}
The Majority Logic OR (MOR) board is another custom-designed VME-style module used to implement RPC layer-coincidence requirements in hardware and supports up to ten input trigger signals in TTL format. 
In the 2024 and 2025 \proanubis configuration, it receives six TTL trigger inputs from the TOR boards, corresponding to the $\eta$ side readout of six RPCs. 
These input signals are summed in an analogue fashion, after which the combined voltage is compared with a configurable threshold using an on-board comparator.

The threshold level is adjustable via an external remotely operable power supply, which can be used to apply the reference voltage to a comparator, internally divided into ten steps, enabling the selection of different coincidence multiplicities. 
In practice, this reference voltage translates into the requirement for $N=1,2,...6$ out of 6 RPC detectors to coincidentally register a signal on the $\eta$ side.
In this way, the MOR board allows the required RPC layer multiplicity to be selected based on the desired level of background suppression.

When the analogue sum exceeds the set threshold, the MOR board issues an output signal in TTL format with a pulse width of 60~ns and a latency of around 10~ns. 
The trigger output is then made available to the remainder of the \proanubis DAQ chain for further processing.

\subsubsection{TTL-to-ECL Signal Converter Board}
\label{SubSec:TTL2ECL}
A custom-designed TTL-to-ECL VME-style signal converter board provides protocol translation required for communication with the TDCs, which only accept ECL-level control inputs. 
The board receives the TTL majority-trigger signal from the MOR board via a LEMO connector, converts it into the ECL format, and delivers the converted signal to the TDCs through an 8-pin flat-cable control interface.

In addition, the board synchronously distributes the 40~MHz LHC clock, received from the TIM in the ECL format, to the TDC control input via the same connector standard. 
In this way, the converter board ensures that both the global timing reference and the correctly formatted trigger signal are supplied to the TDC front-end, maintaining signal integrity and compatibility within the \proanubis DAQ chain.

\subsection{Secondary DAQ rack} 
\label{proAnubis-DAQ1}
In addition to the primary DCS and DAQ rack located in the experimental cavern, a secondary DAQ rack has been installed in USA15, the secure and radiation-protected \atlas server room. 
This rack hosts two backup servers that are connected to the \proanubis DAQ network via a dedicated switch, enabling direct access to the VME Controller (Section~\ref{sec:CAENcontroller}) and full control of the DAQ system. 
These servers can assume the role of the primary DAQ machines if required, providing identical data acquisition and control functionality.

The motivation for this redundancy arises from the higher radiation levels in the cavern environment, which can gradually degrade or, in some cases, cause sudden failure of the primary server hardware during LHC operation, potentially leading to a loss of communication and control to \proanubis setup. 
By maintaining additional operational backup servers in USA15, \proanubis ensures better control and data acquisition capability throughout standard LHC operations, thereby improving system reliability and operational resilience.
Retrospectively, backup servers never had to be engaged.

\section{Signal Processing and Data Acquisition} 
\label{Sec:Signal_processing}
The signal processing and data acquisition chain of the \proanubis detector is designed to extract space-time information from RPC signals produced by particle interactions in the \atlas cavern. 
Each RPC detector measures $99\,{\rm cm}\times182\,{\rm cm}$ and consists of two readout planes: the \(\eta\)-plane, which has 32 strips running parallel to the long side, and the \(\phi\)-plane with 64 strips running orthogonally. 
Both planes use copper pickup strips implemented on a Printed Circuit Board (PCB) with a pitch of approximately \(25~\text{mm}\).

The RPC readout is provided by upgraded on-detector FE electronics developed by the \atlas Muon Collaboration~\cite{Pizzimento:2019slz,PIZZIMENTO2024169892}. 
Each RPC is instrumented with 12 FE boards: four on the \(\eta\)-plane and eight on the \(\phi\)-plane. 
Each FE board reads eight channels, each of which corresponding to a strip. 
Thus, the full \proanubis system comprises:
\[
6~\text{RPCs} \times 12~\text{boards} \times 8~\text{channels}
= 576~\text{readout channels},
\]
of which 192 correspond to \(\eta\) strips and 384 to \(\phi\) strips.

\begin{figure*}[ht]
\centering
\includegraphics[width=12 cm]{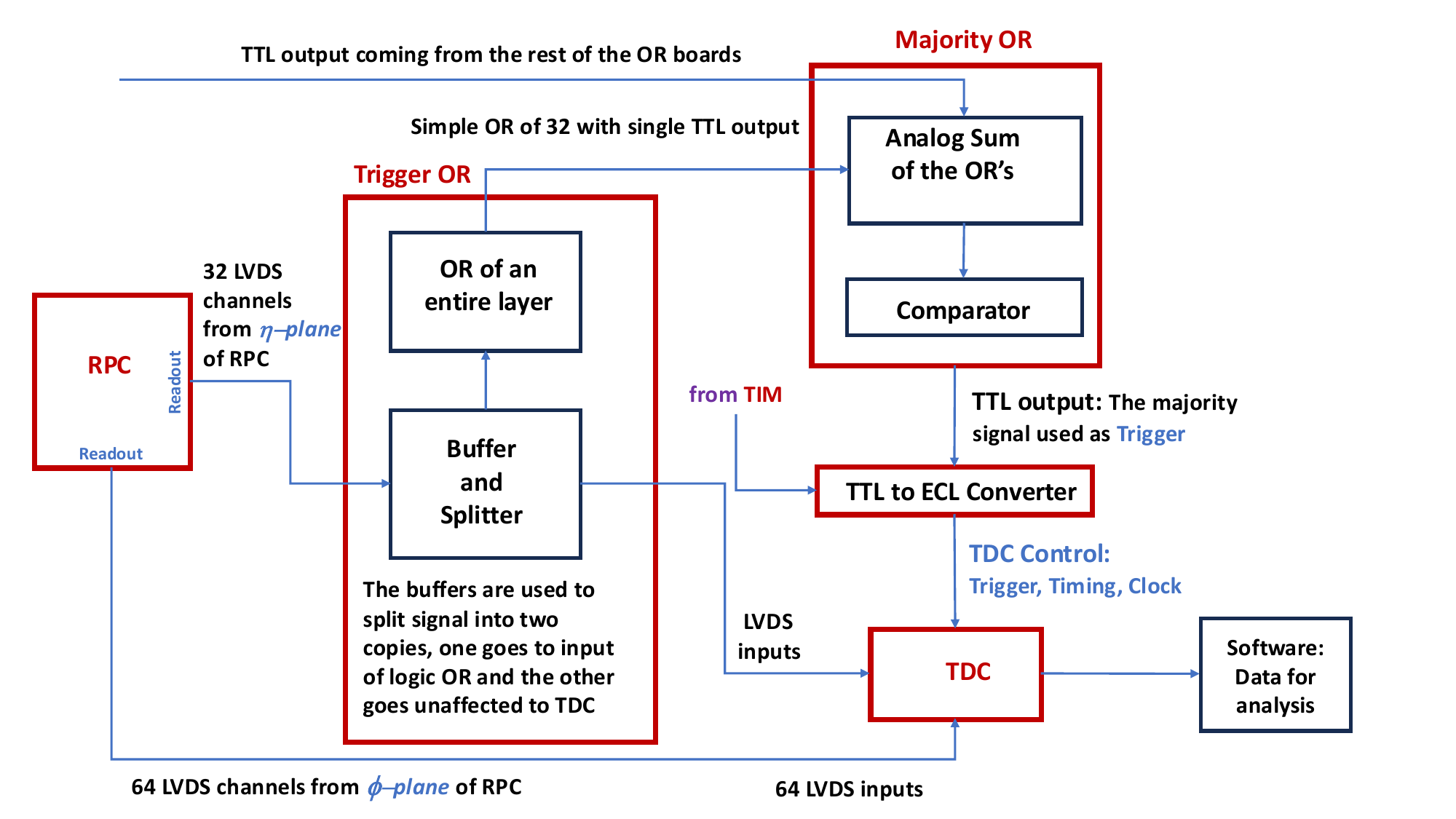}  
\caption{
Schematic overview of the signal flow from the RPC readout to the final data acquisition stage. 
For clarity, the diagram shows only a subset of channels, which includes  32 \(\eta\) and 64 \(\phi\) channels of a single RPC detector; the same architecture applies to each of the six RPC detectors.
} 
\label{fig:proANUBIS_signal_processing}
\end{figure*}

A schematic overview of the signal flow for the 32~\(\eta\) and 64~\(\phi\) channels from a single RPC detector is shown in Figure~\ref{fig:proANUBIS_signal_processing}. 
A charge avalanche induced by a traversing charged particle (\eg a muon) in the gas gap generates signals on the pickup strips, which are amplified and shaped by the FE boards, that are generically labelled as `Readout' in the Figure.
The resulting differential LVDS pulses from the \(\phi\)-readout plane are routed directly to the TDCs, where their arrival times are digitised. 
The same direct routing applies to all \(\phi\) channels in the system.

For the \(\eta\)-readout planes, an additional step is performed before the readout signals are input to TDCs, which implements a hardware-trigger path. 
The LVDS signals from FE boards are routed to the TOR board~(Section~\ref{SubSec:TOR}), which buffers and splits the signals into two paths. 
One path simply channels the original LVDS signals to the TDCs in pass-through mode. 
The other path generates a TTL trigger signal whenever one or more strips fire in a 25 ns time window. 
The trigger is implemented on the $\eta$ side only since it covers the entire active RPC area with only 32 channels. 
Given the high efficiency of RPC detectors, the additional benefit of triggering on the $\phi$ side in addition to the $\eta$ side would be marginal.

The TTL trigger signals from the six TOR boards, each corresponding to an RPC detector, are passed to the MOR board~(Section~\ref{SubSec:SOR}), where a configurable coincidence requirement is applied to suppress uncorrelated background signals.
Upon meeting the set $N$ out of 6 condition, the MOR board generates the main trigger signal with a width that is set to 60~ns.

Subsequently, the trigger output is translated into the ECL format by a custom converter board~(Section~\ref{SubSec:TTL2ECL}) to satisfy the TDC control-input specifications. 
The same board distributes the 40~MHz LHC clock received from the TIM~(Section~\ref{SubSec:TIM}) to the TDCs, providing the low-jitter timing reference needed for synchronisation with ATLAS and ultimately the LHC.

Finally, the digitised data are buffered on the CAEN V767 TDCs~(Section~\ref{sec:CAENTDCs}) and retrieved through the V4718 VME–Ethernet Controller~(Section~\ref{sec:CAENcontroller}). 
The DAQ supports both triggerless and triggered readout modes. 
Five TDCs operating in triggered mode record per-event hit times from RPCs, while a sixth TDC running in triggerless mode continuously records copies of the trigger signal from \proanubis setup and LHC bunch-crossing identifier (BCID) from the TIM, enabling offline time alignment with \atlas collision events.
A custom C/C++ data acquisition software called OSIRIS (OperationS and monItoRIng Software) performs data decoding, event building, and real-time monitoring.

\section{Remote Control and Operation Capabilities} 
\label{Sec:Remote_Control}
The \proanubis detector and its DAQ system are designed for fully remote operation, as physical access to the \atlas cavern is highly restricted during data-taking periods. 
This section outlines the key components that enable flexible and continuous control of the setup, ensuring that essential actions such as power cycling, system resets, or monitoring adjustments can be performed safely and efficiently without on-site intervention.

\subsection{Power Distribution Unit} 
\label{sec:PowerDistributionUnit}
The PDU~\cite{Ref_ANUBIS_PDU}, shown in Figure~\ref{fig:proANUBIS_PDU_plus_CANbus}, distributes electrical power to the \proanubis detector and all components housed in the primary DCS and DAQ rack. 
With 12 individually switchable power output ports, it supplies the computer servers, the network switch, the environmental monitoring system (Section~\ref{Sec:Ambient_monitoring}), the VME crates, and the CAEN SY4527 mainframe which then provides HV and LV to the RPCs detectors and DAQ electronics.

The PDU is accessible through either a secure web interface or command-line utilities, allowing remote control of all components powered by it. 
This functionality is particularly useful for system recovery in the event of unexpected interruptions, as individual devices can be power-cycled or reset without requiring physical access to the DCS and DAQ rack.

A watchdog function implemented in the PDU  provides an additional layer of operational control and protection. 
For example, the network switch is self-monitored continuously: should it become unreachable, the PDU automatically detects the loss of communication and initiates a reset of the corresponding port, restoring connectivity.

\subsection{CANbus System for VME Crates} 
\label{sec:CANbus}
\begin{figure}[ht]
\centering
\begin{subfigure}{\linewidth}
\centering
\includegraphics[width=6.5 cm]{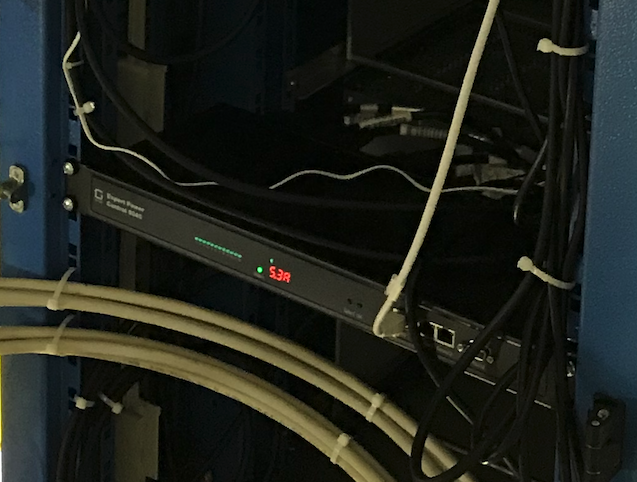} 
\caption{}
  \label{fig:proANUBIS_PDU_plus_CANbus_Top}
\end{subfigure}
\begin{subfigure}{\linewidth}
\centering
\includegraphics[width=6.5 cm]{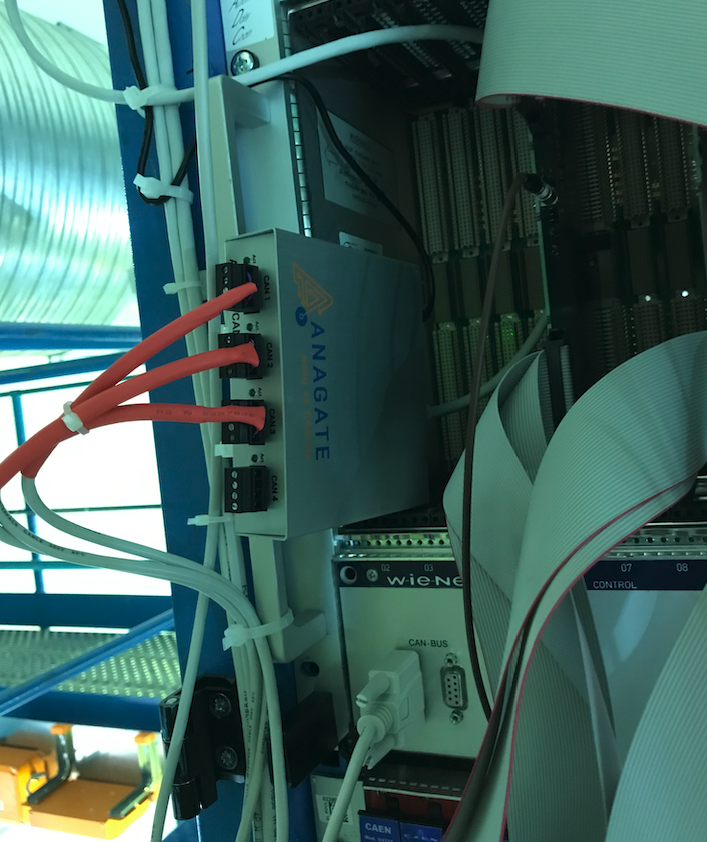}
\caption{}
  \label{fig:proANUBIS_PDU_plus_CANbus_Low}
\end{subfigure}
\caption{
(a) The Power Distribution Unit installed on the backside of the primary DCS and DAQ rack, providing power to the major system elements through individually controlled channels. 
(b) The AnaGate CAN F4 interface used to remotely control the power state of the two VME crates. 
Three active CAN channels are identified by the red cables.}
\label{fig:proANUBIS_PDU_plus_CANbus}
\end{figure}

Remote power control of the VME crates is achieved via a Controller Area Network bus (CANbus) interface. 
Communication between the DAQ servers and the VME crate power controllers is established through the AnaGate CAN Gateway F4 device~\cite{Ref_ANUBIS_AnaGate}, shown in Figure~\ref{fig:proANUBIS_PDU_plus_CANbus}. 
The AnaGate unit operates over the standard TCP/IP network and is accessed through an OPC (Open Platform Communication) UaExpert client application~\cite{opc_uaExpert, Ref_ANUBIS_OPC_clients}, which provides a graphical interface for interacting with the crate control logic. 
Once the OPC client is connected to the gateway, users can remotely execute commands such as power-cycling individual VME crates, enabling quick intervention without requiring physical access to the ATLAS experimental cavern.

\subsection{HV, LV and Server Power Management}
\label{sec:HVLVcontrol}
Remote control of the HV and LV systems is provided through the single-board computer implemented on the CAEN SY4527 mainframe, which allows remote ramping, shutdown, and channel-by-channel configuration of the HV channels for the detector and LV channels for the Front-End electronics and the hardware trigger. 
In parallel, the computer servers can be power-cycled remotely either via the PDU or through Wake-on-LAN commands that trigger MAC-address–targeted resets. 
This dual mechanism provides an additional layer of operational redundancy, enabling recovery from software crashes or network failures without requiring physical access to the cavern.

\section{Monitoring of Ambient and Gas Conditions} 
\label{Sec:Ambient_monitoring}
Accurate environmental information such as the ambient temperature, pressure, and relative humidity is essential for stable RPC operation, as these parameters influence the gas ionisation properties and hence the detector gain. 
To ensure that environmental parameters are properly accounted for in performance studies, a cost-effective `weather station' was developed using readily available commercial components. 
The system incorporates two complementary sensor technologies. 

The first sensor, a BME280 module~\cite{sensor_boschbme280}, measures the ambient temperature, pressure, and relative humidity close to the detector. 
These measurements provide the information needed to evaluate atmospheric variations in the cavern and their potential influence on the detector response.

\begin{figure}[htb]
\centering
\includegraphics[width=9cm]{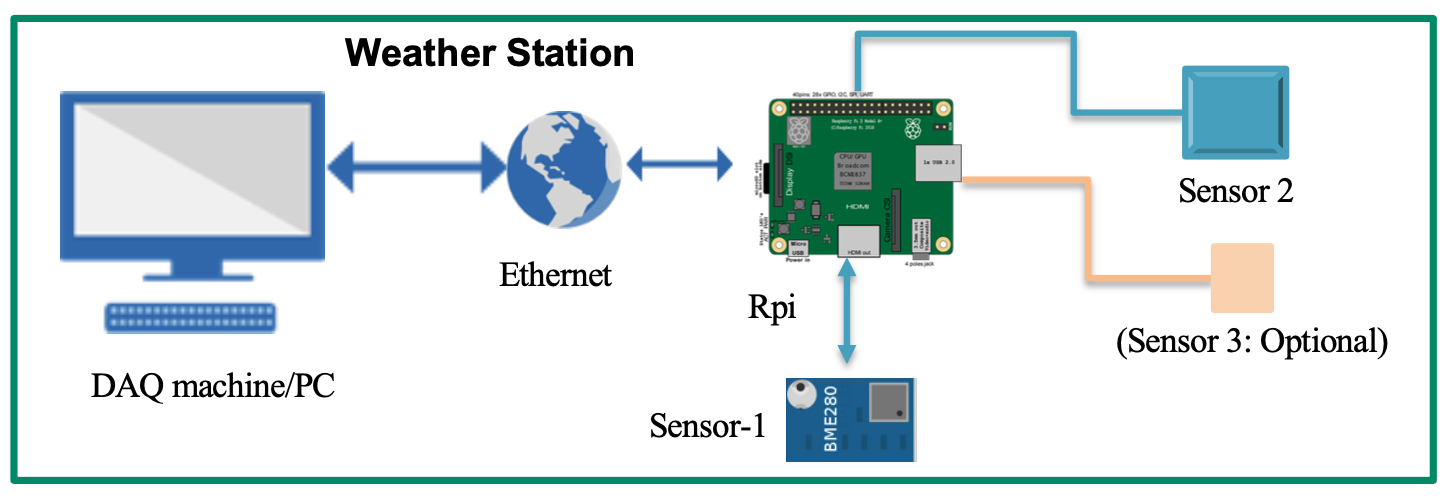} 
\caption{Schematic of the environmental monitoring system where a Raspberry Pi (labelled as Rpi) interfaces with two sensors, BME280 (ambient conditions) and SHT85 (gas conditions), enabling remote monitoring of conditions surrounding \proanubis. 
}
\label{fig:proANUBIS_WS_station_Rpi}
\end{figure}

\begin{figure}[htb]
\centering
\begin{subfigure}{0.49\linewidth}
\centering
\includegraphics[height=3.5cm]{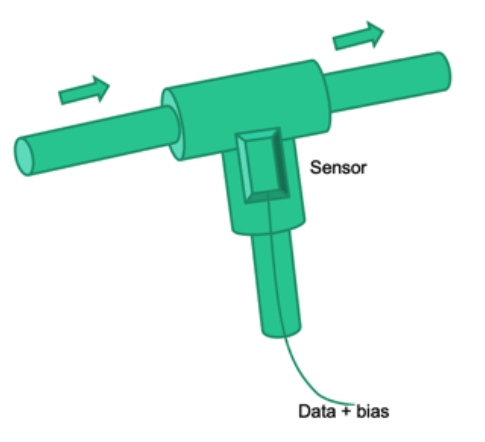}
\caption{}
  \label{fig:proANUBIS_WS_station_Top}
\end{subfigure}
\begin{subfigure}{0.49\linewidth}
\centering
\includegraphics[height=3.5cm]{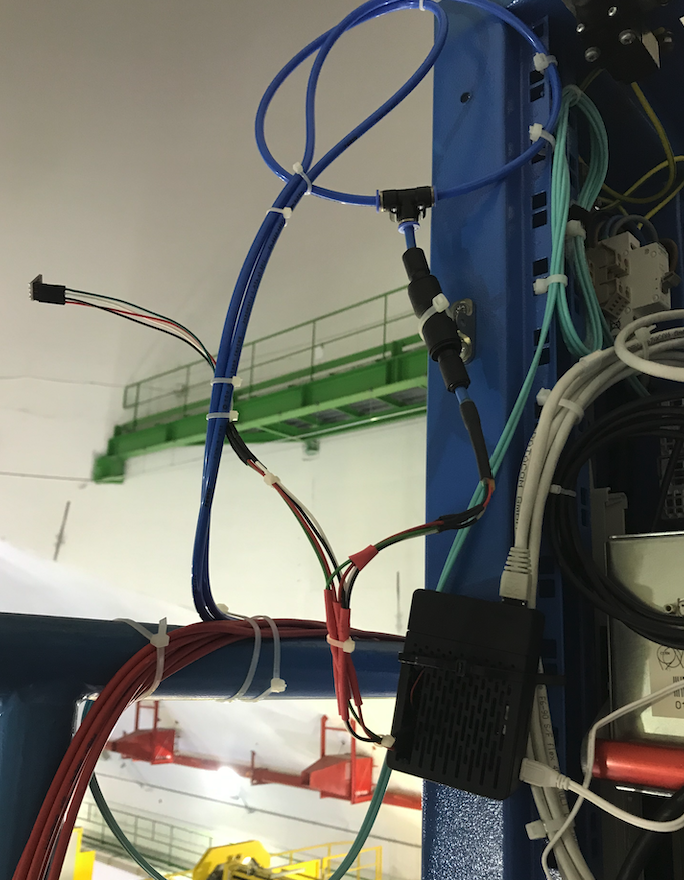}  
\caption{}
  \label{fig:proANUBIS_WS_station_Low}
\end{subfigure}
\caption{ 
(a) Schematic of the SHT85 sensor inside the gas line to monitor the conditions of the RPC gas mixture. 
(b) Installed configuration on the \proanubis setup.
}
\label{fig:proANUBIS_WS_station}
\end{figure}

For monitoring the conditions of the active gas fed into the \proanubis RPC system, an SHT85 sensor~\cite{sensor_sht85} has been installed inside a gas line upstream of the chambers (Figures~\ref{fig:proANUBIS_WS_station_Rpi} and~\ref{fig:proANUBIS_WS_station}). 
Its compact design has allowed in-line deployment, enabling tracking of the conditions of the gas mixture being flushed through the RPCs.

\begin{figure}[ht]
\centering
\includegraphics[width=9 cm, height=5cm]{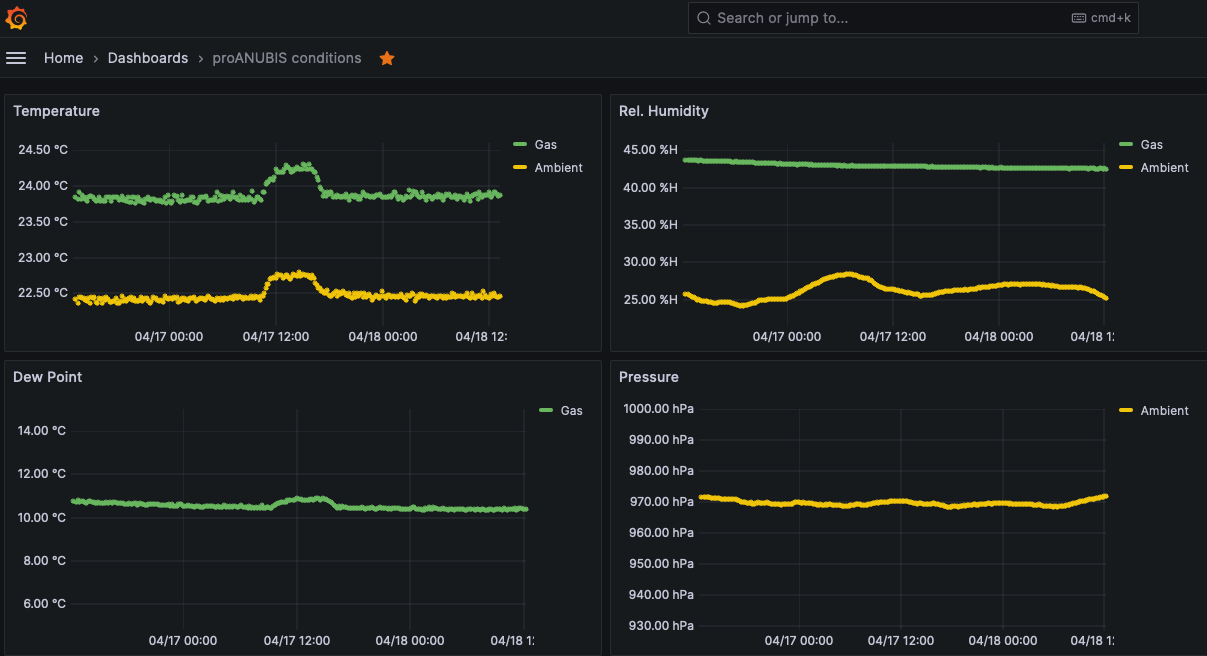}  
\caption{Grafana panel displaying real-time ambient and gas conditions (temperature, relative humidity, and pressure) relevant for \proanubis performance monitoring over a period of two days.}
\label{fig:proANUBIS_grafana_display}
\end{figure}

Both sensors are interfaced through USB ports with a Raspberry Pi 4 Model B, which in turn is interfaced to the DCS system via a TCP/IP network connection, enabling remote control and continuous data acquisition. 
Custom scripts automate data readout while ensuring uninterrupted monitoring during LHC operations. 
All monitored parameters are recorded in an InfluxDB time-series database and visualised in real time through Grafana dashboards, as exemplified in Figure~\ref{fig:proANUBIS_grafana_display}.

\section{First Look at Commissioning Data}
\label{sec:CommissioningData}
The bottom RPC module pair (triplet–singlet) of \proanubis is separated by 60~cm, while the top pair (singlet-doublet) is spaced 50~cm apart. 
This provides a total lever arm of 110~cm for tracking.
The resulting time-of-flight discrimination corresponds to a time difference of
\[
\Delta t \approx \frac{1.1~\metre}{c} \simeq 4~\text{ns},
\]
which is well resolved by the 0.8~ns TDC sampling. 
Cosmic-ray muons predominantly traverse the system from top to bottom, while particles from $pp$ collisions produced at the \atlas interaction point travel upward. 
The ordering of hit times across the three planes therefore enables an event-by-event determination of the particle direction with a high confidence.

\begin{figure}[h]
\flushleft
\includegraphics[width=8.6cm]{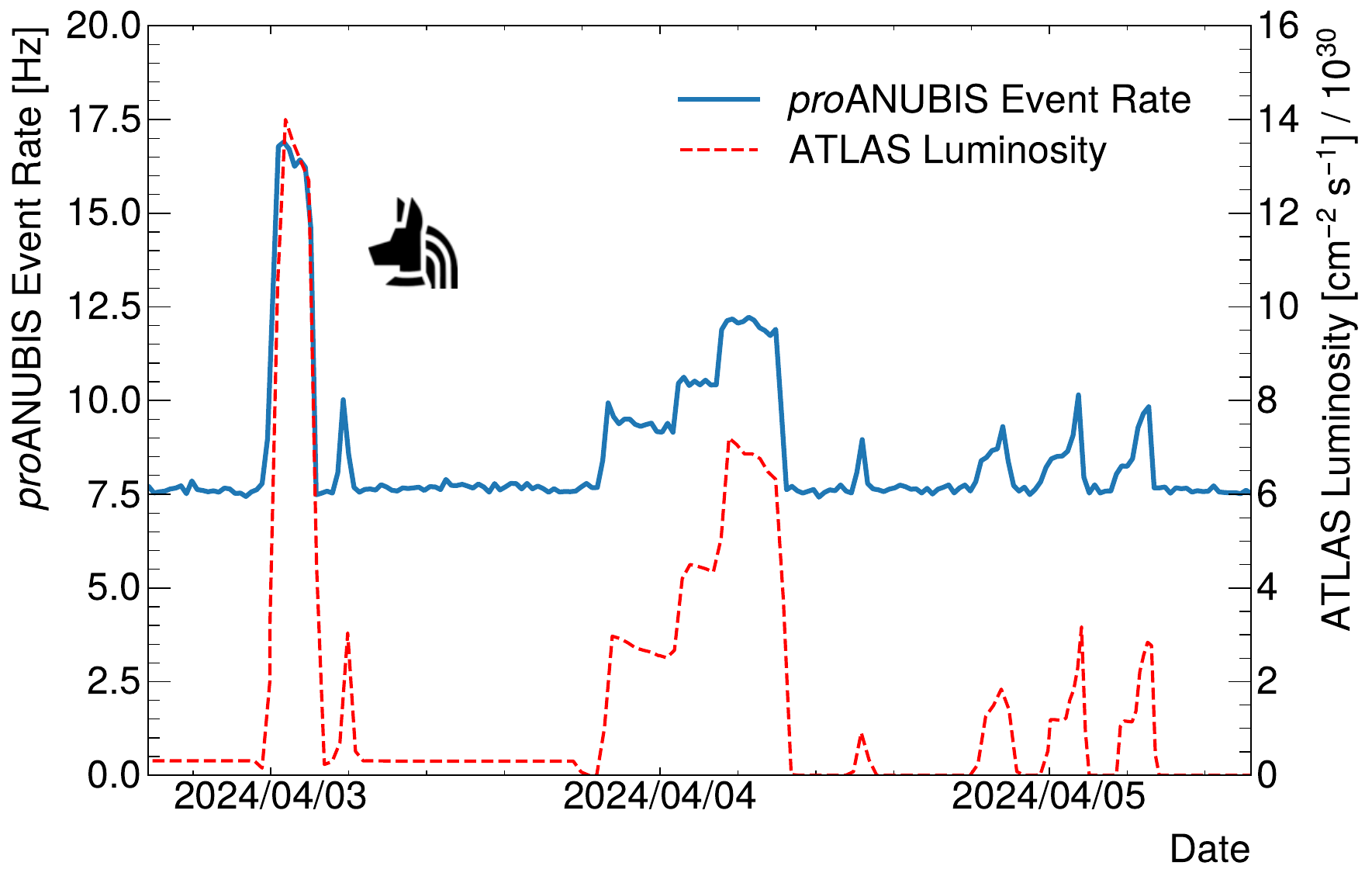}  
\caption{First event rates recorded by the \proanubis detector during a $\sim$three day LHC beam commissioning period, showing a clear correlation between the \proanubis trigger rate and the instantaneous luminosity measured by the \atlas experiment.} 
\label{fig:proANUBIS_lumi_display}
\end{figure}

Early data taking indicates a cosmic muon rate of approximately 7--9~Hz using a 3 out of 6 trigger logic; for instance, in Figure~\ref{fig:proANUBIS_lumi_display} before the first luminosity peak during LHC operation, a rate of about 7.5~Hz is observed. 
This value is consistent with expectations for a detector located roughly 80~\metre underground and reflects the higher likelihood of particle detection within the RPC triplet, even for oblique muon trajectories. 
This is because the triplet RPC module can fulfil the 3 out of 6 trigger condition just by itself, maximising angular acceptance.
Moreover, the detector is situated close to the edge of the \atlas PX14 access shaft (see  Figure~\ref{fig:proANUBIS_installation_position}) and hence is only partially covered by overburden material such as concrete and rock. 
As a result, cosmic ray muons entering through this shaft can relatively easily cross the \proanubis detector volume, hence producing a pronounced accumulation in the observed angular distribution. 

\begin{figure}[ht]
\centering
\includegraphics[width=8.25cm]{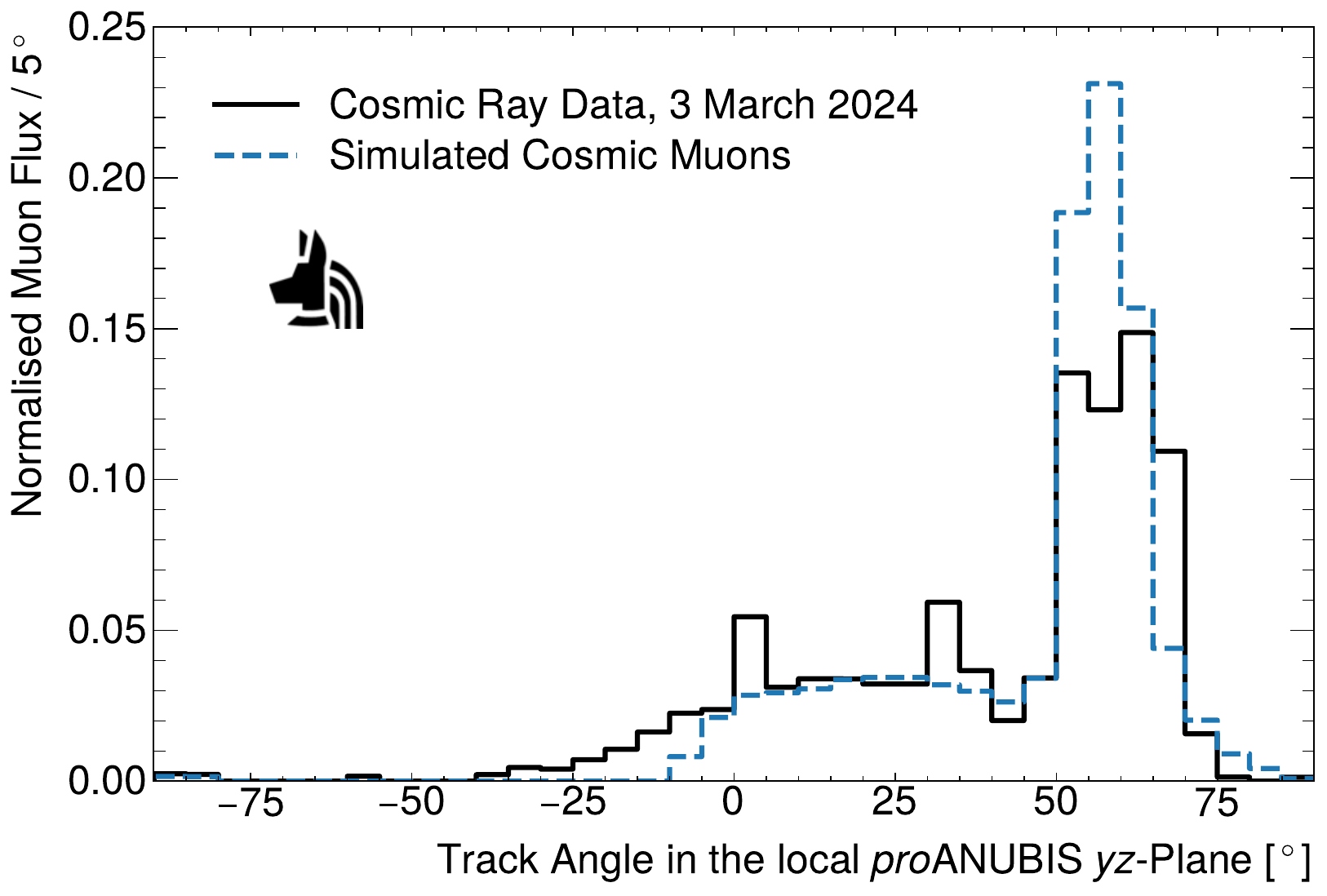}  
\caption{
Measured angular distribution of tracks from cosmic ray muon candidates in the local $yz$-plane of the \proanubis detector, compared with simulation. 
} 
\label{fig:proANUBIS_simcosmics}
\end{figure}

Figure~\ref{fig:proANUBIS_simcosmics} shows the measured angular distribution of tracks from cosmic ray muon candidates in the local $yz$-plane\footnote{The local coordinate system of \proanubis is centred at the bottom left corner of the lowest RPC detector in the triplet RPC module when seen from the other RPC modules. The $x$ axis is aligned with the long side of the RPC detector, the $y$ axis with its short side, and the $z$ axis is perpendicular to the detector plane, forming a right-handed coordinate system. The line of sight from \proanubis to the IP approximately coincides with the $-z$ axis.} of the \proanubis detector. 
The observed higher flux near the shaft direction at around $60^\circ$ corresponds to the reduced overburden in that region, while the two smaller peaks at $0^\circ$ and $30^\circ$ degrees are attributed to the finite strip granularity of the \proanubis RPC detectors impacting tracks registered only by the triplet RPC module. 
Apart from these localised features, the data follow a smooth angular dependence consistent with expectations for an underground detector exposed to the attenuated flux of cosmic ray particles.

The measured distribution is compared with a simple MC simulation based on measured fluxes and angular distributions for cosmic ray muons of various energy ranges at sea level~\cite{ACHARD200415,MOTOKI2003113}, along with simulated muon survival probabilities based on the rock overburden above the ATLAS cavern determined with \textsc{Geant4}\xspace~\cite{GEANT4:2002zbu}. 
The overall agreement between data and simulation is satisfactory within the statistical and systematic precision of the simulation model.

When a stricter trigger multiplicity condition of 4 out of 6 is imposed, the observed cosmic muon rate decreases to about 1~Hz. 
This reduction arises primarily from the detector geometry: the tracking planes are inclined by $45^{\circ}$ relative to the LHC beam axis and hence are approximately perpendicular to the line of sight towards the IP as seen from the \proanubis location. 
Therefore, only a small subset of downward-going cosmic rays can intersect four out of the six RPC detectors. Furthermore, the singlet and doublet RPC modules are located in regions more heavily shielded by the surrounding rock and infrastructure, further suppressing the multi-layer coincidence rate.

Beyond the cosmic-ray component, a clear correlation between the \proanubis trigger rate and the instantaneous luminosity measured by \atlas, shown in Figure~\ref{fig:proANUBIS_lumi_display}, demonstrates the capability of the \proanubis detector to detect particles originating from $pp$ collision events. 
This observation validates the functionality of the trigger logic and overall DAQ performance, confirming that \proanubis can effectively record both cosmic-ray and collision-induced activity within the UX1 \atlas experimental cavern.

\section{Summary and outlook} 
\label{Sec:Summary}
The ANUBIS experiment, proposed to significantly expand the sensitivity of existing and approved experiments to neutral LLPs by instrumenting the \atlas underground cavern with tracking detectors, has reached a major milestone with the successful installation and commissioning of its prototype detector, \proanubis, in 2023-24. 
Observations of collision events correlated with \atlas data during LHC beam periods, together with measurements of cosmic-ray rates being consistent with expectations, confirmed the correct operation of the detector and trigger chain. 

The \proanubis detector provides a unique opportunity to obtain \textit{in-situ} measurements of the expected background flux utilising background-enriched kinematic regions in a location that is representative for the future ANUBIS detector. 
This allows for refining reconstruction algorithms and related selections, background simulations, and further optimisations.
The data recorded by the \proanubis detector in 2024 and 2025 allowed for a full characterisation of the detector performance considering parameters like efficiency, timing performance, and spatial resolution. 
Studies of background rates and event composition are ongoing and first physics results based on LHC Run~3 data will be reported in forthcoming publications.
The operational experience and physics insight gained from the prototype detector will directly support the transition toward a full-scale installation in the \atlas cavern, while significantly reducing the technical and physics risks associated with deploying a large-scale LLP detector without prior \textit{in-situ} validation.


The broader ANUBIS programme has gained significant momentum on both the simulation and hardware fronts, as highlighted in the submission to the ESPPU submission~\cite{PBC:2025sny, Brandt:2025fdj}.
Recent simulation studies~\cite{ANUBIS:2025sgg} demonstrate a substantial increase in sensitivity to LLP decays compared to existing \atlas subsystems and other transverse proposals~\cite{Gligorov:2017nwh, aitken2025conceptualmathusla} with ongoing efforts extending these projections across a variety of benchmark models. 
R\&D initiatives are underway to search for a new low-Global Warming Potential (GWP) gas mixtures for the replacement of existing RPCs high-GWP gas mixtures, supporting the long-term goal of an environmentally sustainable ANUBIS detector system.

The ANUBIS collaboration continues to grow, with dedicated working groups being established for simulation, detector R\&D, and integration planning. 
With the successful operation of \proanubis and the advancing simulation and design studies, the ANUBIS programme is strongly positioned to progress towards a full implementation within the \atlas cavern in the coming years, pending endorsement by the \atlas Collaboration, the LHCC, and CERN management.

\section*{Acknowledgements}
We thank CERN for the successful operation of the LHC. 
We thank the \atlas collaboration for facilitating the installation of the \proanubis setup in the \atlas underground cavern, and would like to highlight the invaluable technical support from the \atlas Technical Coordination team in particular. 
We would like to thank the CODEX-b Collaboration for lending readily available FE boards, which allowed for a swift construction and installation of \proanubis. 
We thank the technical support teams at the CERN BB5 laboratory, in particular Luigi Di Stante, Spyridon Kompogiannis, Markus Lippert, and Enrico Tusi for their advice with the assembly and integration of the \proanubis detector. 


\printbibliography[title=References,heading=bibintoc]

\end{document}